\documentclass[aps,pra,twocolumn,showpacs]{revtex4}
\usepackage{amsmath,amssymb,graphicx,bbold}
\usepackage{graphicx}  
\usepackage{dcolumn}   
\usepackage{bm}        
\usepackage{color}

\providecommand{\abs}[1]{\left\lvert#1\right\rvert}

\begin{document}

\title{Orbital angular momentum mixing in type II second harmonic generation}

\vskip \baselineskip

\author{Leonardo J. Pereira}

\author{Wagner T. Buono}

\author{Daniel S. Tasca}

\author{Kaled Dechoum}

\author{Antonio Z. Khoury}

\affiliation{Instituto de F\'{\i}sica - Universidade Federal Fluminense \\
Av. Gal. Milton Tavares de Souza S/N, 24210-340 Niter\'oi - RJ, Brazil}

\date{\today}

\begin{abstract}
We investigate the non linear mixing of orbital angular momentum in type II second harmonic generation 
with arbitrary topological charges imprinted on two orthogonally polarized beams. Starting from the 
basic nonlinear equations for the interacting fields, we derive the selection rules determining 
the set of paraxial modes taking part in the interaction. Conservation of orbital angular momentum 
naturally appears as the topological charge selection rule. However, a less intuitive rule applies to 
the radial orders when modes carrying opposite helicities are combined in the nonlinear crystal, an 
intriguing feature confirmed by experimental measurements.
\end{abstract}
\maketitle

\vskip \baselineskip

\section{Introduction}

The ability to control different degrees of freedom of a light beam is essential for both quantum and 
classical communication protocols. In this context, orbital angular momentum (OAM) has proved to be a potentially 
useful tool and has motivated a fair amount of research work with potential applications to 
quantum and classical information processing \cite{oam25}. 
In the quantum domain, qubits and qudits can be encoded on Laguerre-Gaussian (LG) or Hermite-Gaussian (HG) modes 
that combined with the photon polarization allows creation of entanglement between internal photonic degrees of 
freedom. Many works have been devoted to schemes for implementing and applying this spin-orbit coupling 
\cite{poincare,imagepol,topoluff,quplate,simon,aiello1,cardano,milione,aiello2,structured-waves}. 
Beyond the intrinsic beauty of this subject, one may find interesting applications to quantum information 
tasks like optical communication \cite{milione2,network}, teleportation schemes 
\cite{chen2009,barreiro2010,khourymilman,rafsanjani,remote-state-boyd,szameit,teleportUFF}, alignment free 
quantum cryptography \cite{cryptosteve,cryptouff,cryptolorenzo}, controlled gates for quantum computation 
\cite{cnotsteve,cnotuff}, quantum simulations \cite{josabuff,environ} and metrology \cite{aiello3,forbes,forbes3}. 
Quite curiously, the presence of spin-orbit structures in an optical beam can be characterized by inequality 
criteria similar to those used to witness entanglement in quantum mechanics 
\cite{bell2,chen,bellpadgett,kagalwala,eberly,eberly2,tripartite}.  

Orbital angular momentum exchange in nonlinear interactions has been extensively studied and is still a fruitful 
domain. It has already been investigated in frequency up 
\cite{SHG,SUM,OAM-SHG-Li,OAM-SHG-UFF,sum-freq,algebra,twisted-nonlinear,beam-prop,shg-plasmonic}, 
and down \cite{PDC1,zeilinger,PDCUFF} 
conversion, optical parametric oscillation \cite{Schwob,OPO1,OPOUFF,OPO2}, four-wave mixing in atomic vapors \cite{ATO1,ATO2} 
and high harmonic generation \cite{atto,high-hg}. The control of the nonlinear optical interaction through 
polarization has been also considered both in the quantum \cite{imagepol} and classical regimes 
\cite{OAM-SHG-UFF,pol-shaping}.  
In these examples, the phase match condition imposes simple arithmetic relations among the topological charges 
of the interacting modes \cite{OAM-SHG-UFF,algebra}. 
In the seminal works on second harmonic generation with OAM beams, the intuitive charge 
doubling condition was demonstrated and remained as a natural assumption until other degrees of freedom were 
brought into play. In type II second harmonic generation with LG modes, we have demonstrated different topological 
charge operations controlled by the polarizations of the interacting beams. Actually, these 
operations are a natural consequence of selection rules arising from the spatial overlap among the interacting modes. 
In cylindrical coordinates, the angular part of the spatial overlap trivially leads to topological charge conservation. 
There are also more subtle selection rules arising from the radial part \cite{radial-pdc}. 

In this work we perform a 
detailed study about orbital angular momentum mixing in collinear type II second harmonic generation. 
Multimode coupling is considered in connection with OAM addition and the selection rules are derived 
based on the spatial overlap between the modes participating in the nonlinear process. 
One curious consequence of these selection rules is the appearance of higher radial orders 
when opposite topological charges are added. This intriguing feature of the nonlinear interaction is 
theoretically derived and experimentally confirmed. We also derive the analytical solution for the nonlinear 
dynamical equations both for co- and counter-rotating OAM modes, as a multimode generalization of the result 
obtained in \cite{bloembergen}. Another curious feature of the nonlinear 
coupling is the reduction of the apparent multimode to an effective three-mode dynamics when counter 
rotating vortices are mixed. The dynamics of the higher radial modes is \textit{slaved} by the 
fundamental one, making an effective three-mode coupling.

\section{Dynamical equations for the coupling modes}

We will consider type II second harmonic generation in collinear configuration with orthogonally polarized 
incoming beams carrying arbitrary topological charges. 
Let us start with the expression for the electric field vector corresponding to a light wave with the 
fundamental frequency $\omega\,$, propagating along the $z$ direction
\begin{equation}
\mathbf{E}_\omega (\mathbf{r},z;t) = \left[ \mathcal{E}_h(\mathbf{r},z)\,\mathbf{\hat{e}}_{h}\, e^{i k_{h}z} 
+ \mathcal{E}_v (\mathbf{r},z)\,\mathbf{\hat{e}}_{v}\,e^{i k_{v}z} \right]
e^{-i\,\omega t}\;,
\end{equation}
where $\mathbf{\hat{e}}_h$ ($\mathbf{\hat{e}}_v$), $k_h$ ($k_v$) and $\mathcal{E}_h$ ($\mathcal{E}_v$) are, respectively, the polarization unit vector 
along the horizontal (vertical) direction, the corresponding wave number and transverse spatial function. 
For type II phase match the second harmonic field can be written as
\begin{equation}
\mathbf{E}_{2\omega} (\mathbf{r},z;t) = \mathcal{E}_{2\omega}(\mathbf{r},z)\,\mathbf{\hat{e}}_{h}\,e^{i \left ( k_{2\omega} z - 2\omega t \right )}\;.
\end{equation}
The three field components, $(\omega,h),(\omega,v)$ and $(2\omega)\,$, follow a coupled evolution inside 
the nonlinear medium with a coupling constant $\chi$ that involves the relevant terms of the nonlinear susceptibility tensor. 
Under phase match ($k_{2\omega} = k_{h} + k_{v}$), the equations describing this coupled evolution in the paraxial approximation are
\begin{eqnarray} 
\nabla_\perp^2 \mathcal{E}_{h} + 2ik_{h}\frac{\partial \mathcal{E} _{h}}{\partial z} &=& -\frac{\chi\,\omega^2}{c^2}\,\mathcal{E} _{2\omega}\,\mathcal{E} _{v}^*\;, 
\nonumber \\
\nabla_\perp^2 \mathcal{E}_{v} + 2ik_{v}\frac{\partial \mathcal{E} _{v}}{\partial z} &=& -\frac{\chi\,\omega^2}{c^2}\,\mathcal{E} _{2\omega}\,\mathcal{E} _{h}^*\;, 
\\
\nabla_\perp^2 \mathcal{E}_{2\omega} + 2ik_{2\omega}\frac{\partial \mathcal{E} _{2\omega }}{\partial z} &=& -4\,\frac{\chi\,\omega^2}{c^2} \mathcal{E} _{h}\,\mathcal{E} _{v}\;,
\nonumber
\end{eqnarray}
with $\nabla_\perp\equiv(\partial/\partial x, \partial/\partial y)\,$.
In order to inspect the OAM exchange in the nonlinear process, we expand the interacting fields in the Laguerre-Gauss basis
\begin{eqnarray}
\mathcal{E} _{h}\left(\mathbf{r},z\right) &=& \sqrt{\frac{\omega}{n_h}}\,\sum_{p, l} A_{p l}^h\,u_{p l}^h \left ( \mathbf{r},z \right )\;,
\nonumber \\
\mathcal{E} _{v}\left(\mathbf{r},z\right) &=& \sqrt{\frac{\omega}{n_v}}\,\sum_{p, l} A_{p l}^v\,u_{p l}^v \left ( \mathbf{r},z \right )\;,
\\
\mathcal{E}_{2\omega}\left(\mathbf{r},z\right) &=& 
\sqrt{\frac{2\omega}{n_{2\omega}}}\,\sum_{p, l} B_{p l}\,u_{p l}^{2\omega}\left ( \mathbf{r},z \right )\;, 
\nonumber 
\end{eqnarray}
where $n_{h(v)}$ is the refraction index for the horizontally (vertically) polarized infrared beam and $n_{2\omega}$ is the 
the second harmonic refraction index.
The orthonormal Laguerre-Gaussian mode functions are
\begin{eqnarray}
u_{p l}^j (\mathbf{r},z) &=& \sqrt{\frac{2}{\pi}}\,\frac{\mathcal{N}_{p\,l}}{w_{j}(z)}\,
\left (\frac{\sqrt{2}\,r}{{w_j} (z)}\right )^{\left | l \right |} L_p^{\left | l \right |}\left ( \frac{2 r^2}{w_j^{\,2} (z)} \right ) 
\\
&\times& e^{-\frac{r^2}{w_j^{\,2} (z)}}\,
e^{i \left [\frac{k_j r^2}{2\,R_j(z)} - 
\left ( 2p + \left | l \right | +1 \right ) \arctan\left ( \frac{z}{z_{R_j}} \right ) + 
l \theta  \right ]}\;, 
\label{uipl}
\nonumber\\
\mathcal{N}_{p\,l} &=& \sqrt{\frac{p!}{\left ( p + \left | l \right | \right )!}}\;,
\end{eqnarray}
$(j=h,v,2\omega)\,$.
The beam width and wavefront radius at position $z$ are respectively given by
\begin{eqnarray}
w_j (z) &=& w_j\,\sqrt{1 + \left ( \frac{z}{z_{R_j}} \right )^2} \;,
\\
R_j(z) &=& z\,\left[ 1 + \left ( \frac{z_{R_j}}{z} \right )^2\right] \;,
\end{eqnarray}
where $w_j = \sqrt{2 z_{R_j}/k_j}$ is the beam waist and $z_{R_j}$ is the Rayleigh distance.
Besides the longitudinal phase match condition ($k_h + k_v = k_{2\omega}$), efficient 
frequency conversion also requires transverse phase match, imposing the wavefront overlap 
$R_h = R_v = R_{2\omega}$ and a common Rayleigh range $z_{R_h} = z_{R_v} = z_{R_{2\omega}} = z_R\,$. 

This transverse mode decomposition allows the description of the field evolution as a set of coupled equations 
for the mode amplitudes \cite{Schwob},
\begin{eqnarray}
\frac{d A_{pl}^h}{d z} &=& i\,g \sum_{p'l'}\sum_{p''l''}\Lambda _{p'pp''}^{l'll''}\,\,B_{p'l'} \left(A_{p''l''}^v\right)^*\;, 
\nonumber \\
\frac{d A_{pl}^v}{d z} &=& i\,g \sum_{p'l'}\sum_{p''l''}\Lambda _{p'p''p}^{l'l''l}\,\,B_{p'l'} \left(A_{p''l''}^h\right)^*\;, 
\label{princ0}\\
\frac{d B_{pl}}{d z} &=& i\,g \sum_{p'l'}\sum_{p''l''}\left(\Lambda _{pp'p''}^{ll'l''}\right)^* \,A_{p'l'}^h A_{p''l''}^v\;.
\nonumber
\end{eqnarray}
The following convenient parameters are introduced
\begin{eqnarray}
g &=& \frac{\chi}{2 c}\,\sqrt{\frac{2\omega^3}{n_h n_v n_{2\omega}}}\,R_{000}^{000}\;, 
\\
\Lambda _{pp'p''}^{ll'l''} &=& \frac{R_{pp'p''}^{ll'l''} }{R_{000}^{000}}\;,
\\
R_{pp'p''}^{ll'l''} &=& \int u_{pl}^{2\omega} \left(u_{p'l'}^h\right)^* \left(u_{p''l''}^v\right)^* d^2\mathbf{r}\;.
\label{Rppp}
\end{eqnarray}
Here $R_{pp'p''}^{ll'l''}$ is the three-mode spatial overlap of Laguerre-Gauss modes with indexes $pl$, $p'l'$ and $p''l''\,$.

Equations (\ref{princ0}) describe the amplitude evolution of each component in the expansion.
We will neglect nonlinear losses ($\chi^*=\chi$) and the Gouy phase acquired inside the crystal, 
making $\left(\Lambda _{pp'p''}^{ll'l''}\right)^* = \Lambda _{pp'p''}^{ll'l''}\,$.


\subsection {Effective nonlinear mode coupling}
\label{Lambdas}

In the multimode dynamics, the effective nonlinear coupling between the different modes is basically ruled by the nonlinear 
susceptibility $\chi$ and the spatial overlap integral defined in Eq. (\ref{Rppp}). Here we are interested in the nonlinear 
OAM mixing of two beams carrying topological charges $l'$ and $l''\,$, with zero radial order ($p'=p''=0$). 
In appendix \ref{appoverlap}, the calculation of the corresponding overlap integral is detailed and two selection rules are 
derived. 
The first one leads to the expected OAM conservation, already discussed in previous works 
\cite{zeilinger,PDCUFF,OAM-SHG-Li,OAM-SHG-UFF,algebra}. 
The second one is less 
obvious and predicts that higher radial orders are generated in the second harmonic field when opposite helicities 
are combined in the nonlinear process. 

When both input topological charges have the same sign ($l'\cdot l'' \geq 0$), the normalized overlap becomes
\begin{eqnarray}
\Lambda _{p00}^{ll'l''} = \delta _{l,l'+l''} \,\left\{
\begin{matrix}
\sqrt{\frac{\xi_h^{\left| l' \right|}\,\xi_v^{\left| l'' \right|}\,\left ( \left | l' \right | + 
\left | l'' \right | \right )!}{\left | l' \right |! \left | l'' \right |! }} & \left(p = 0\right)\;,
\\
0 & \left(p > 0\right)\;,
\end{matrix} 
\right.
\label{lsmaior}
\end{eqnarray}
where $\xi_j\equiv (w_{2\omega}/w_{j})^2\,$. 
This means that the two input modes couple to a single second harmonic one carrying the added topological charge $l=l'+l''$ 
(OAM conservation) and having zero radial order ($p=0$).

A less intuitive situation is produced when the input modes carry topological charges with opposite signs ($l'\cdot l'' < 0$). 
In this case, the resulting overlap integral becomes
\begin{eqnarray}
\Lambda _{p00}^{ll'l''} = \delta _{l,l'+l''} \left\{
\begin{matrix}
\frac{(-1)^p}{\left(P - p\right)!}\,
\sqrt{\frac{\xi_h^{\left| l' \right|}\,\xi_v^{\left| l'' \right|}\left | l' \right |!\,\left | l'' \right |!}{p!\,
		\left(p + \left | l' + l'' \right |\right)!\,}} & \left(p\leq P\right),
\nonumber\\
0 & \left(p > P\right),
\end{matrix} 
\right.
\\
\label{lsmenor}
\end{eqnarray}
where $P=\mathrm{min}(\abs{l'},\abs{l''})\,$, 
so that higher radial orders, up to the minimum value between $\left|l'\right|$ and $\left|l''\right|\,$,
are generated in the second harmonic field.
Therefore, the nonlinear mixing of opposite helicities implies a more complex dynamics with 
more transverse modes taking part in the nonlinear interaction. 

\subsection{Multimode Manley-Rowe relations}

In order to derive multimode conservation laws, it will be useful to define the phase $(\phi^j_{pl},\psi_{pl})$ and 
intensity $(I^j_{pl},J_{pl})$ variables, in terms of which the mode amplitudes are expressed as 
\begin{eqnarray}
A_{pl}^{h(v)} &=& \sqrt{I_{pl}^{h(v)}}\,e^{i \phi_{pl}^{h(v)}}\;,
\nonumber\\
B_{pl} &=& \sqrt{J_{pl}}\,e^{i \psi _{pl}}\;.
\end{eqnarray}
In appendix A we derive the following Manley-Rowe relations
\begin{eqnarray}
\sum_{p''l''} I_{p''l''}^v(z) - \sum_{p'l'} I_{p'l'}^h(z) &=&  \sum_{p''l''} I_{p''l''}^v(0) - \sum_{p'l'} I_{p'l'}^h(0)\;,
\nonumber\\
\sum_{pl} J_{pl}(z) + \sum_{p'l'} I_{p'l'}^{h,v}(z)  &=& \sum_{pl} J_{pl}(0) + \sum_{p'l'} I_{p'l'}^{h,v}(0)\;.
\nonumber\\
\label{MR2}
\end{eqnarray}
These conservation laws are important for the solutions of the dynamical equations. They identify the natural 
integration constants for the dynamical equations.

\section{Nonlinear mixing of co-rotating vortices}
\label{analytical}

We now solve the nonlinear coupled equations for the three-mode interaction in second harmonic 
generation of co-rotating vortices. As we discussed in section \ref{Lambdas}, when two co-rotating vortices with zero radial 
order are mixed in the nonlinear process, a single second harmonic mode will be excited and a three-mode dynamics is realized.
In this case, Eqs. (\ref{princ0}) are significantly simplified and an analytical solution can be found. 
For this purpose we assume only two nonzero initial amplitudes ($l^{\prime}=m$ and $l^{\prime\prime}=n$)
$A_{0m}^h(0) = A_{0n}^v(0) =\sqrt{I_0}$ ($m\cdot n \geq 0$) 
at the crystal entrance, all other modes being empty. According to the overlap integrals given by Eq.~(\ref{lsmaior}), 
a single second harmonic mode with $p=0$ and $l=m+n$ will be excited. The resulting three-mode dynamical equations are
\begin{eqnarray}
	\frac{d A_{0m}^h}{d z} &=& i\,g \Lambda _{000}^{m n}\,\,B_{0} \,A_{0n}^{v\,*}\;, 
	\nonumber \\
	\frac{d A_{0n}^v}{d z} &=& i\,g \Lambda _{000}^{m n}\,\,B_{0} \,A_{0m}^{h\,*}\;, 
	\label{princ1}
	\\
	\frac{d B_{0}}{d z} &=& i\,g \Lambda _{000}^{m n} \,A_{0m}^h A_{0n}^v\;,
	\nonumber
\end{eqnarray}
where $B_{0}$ is the second harmonic amplitude with zero radial order ($p=0$) and we 
omitted the superfluous index $l$ for simplicity. Equations (\ref{princ1}) can be further 
simplified by defining the following rescaled amplitudes
\begin{eqnarray}
b &=& g\,\Lambda_{000}^{m n}\,B_{0}\;, 
\nonumber\\
a_{h} &=& g\,\Lambda_{000}^{m n}\,A_{0 m}^{h}\;, 
\label{rescaled}\\
a_{v} &=& g\,\Lambda_{000}^{m n}\,A_{0 n}^{v}\;, 
\nonumber
\end{eqnarray}
giving 
\begin{eqnarray}
\frac{d a_{h}}{d z} &=& i b\,a_{v}^{*}\;, 
\nonumber\\
\frac{d a_{v}}{d z} &=& i b\,a_{h}^{*}\;,
\label{rescaledprinc1} \\
\frac{d b}{d z} &=& i a_{h}\,a_{v}\;.
\nonumber
\end{eqnarray}
The analytical solution for this system is derived in appendix \ref{3modeanalytical}. 
It resembles the original solution found in Ref.\cite{bloembergen}.
The resulting output intensities are 
\begin{eqnarray}
J_{0}(z) &=& 2 I_{0} \,\tanh^2 \left (g\,\Lambda_{000}^{m n}\,\sqrt{2I_{0}}\,z\right ) \;,
\\
I_{0 n}^v(z) &=& I_{0 m}^h(z) = I_{0} \,\,\mathrm{sech}^2 \left (g\,\Lambda_{000}^{m n}\,\sqrt{2I_{0}}\,z\right )\;,
\nonumber
\end{eqnarray}
where $I_{0 m}^h = \abs{A_{0 m}^h}^2\,$, $I_{0 n}^v = \abs{A_{0 n}^v}^2\,$, $J_{0} = \abs{B_{0}}^2$ and 
$I_0 = I_{0 m}^h(0) = I_{0 n}^v(0)$ is the common intensity of the input modes at the crystal entrance.
Moreover, along the interaction distance $z$ the phase variables evolve under the following phase match condition   
\begin{eqnarray}
\phi_{0 m}^h(z) + \phi_{0 n}^v(z) - \psi_{0}(z) = \left ( 2N + 1 \right )\frac{\pi }{2} \;,
\label{fases}
\end{eqnarray}
with $N\in\mathbb{Z}\,$, as detailed in appendix C.


\section{Higher radial orders generation from counter rotating vortices}

We now investigate the generation of arbitrary radial orders from the orbital angular momentum mixing of 
counter rotating vortices. As we have already mentioned, radial orders up to the minimum absolute value 
of the mixed OAM are generated in the process. In principle, this could imply a complicated multi mode 
dynamics, but we demonstrate that an effective three-mode dynamics can be derived.  
Let us consider the input modes $A_{0 m}^h$ e $A_{0 n}^v$, with $m\cdot n < 0\,$. Without loss of 
generality, we shall assume $\abs{m} < \abs{n}\,$. 
In this case, the dynamical equations are
\begin{eqnarray}
\frac{d A_{0 m}^h}{d z} &=& i g\,A_{0n}^{v\,*}\,\sum_{p=0}^{\abs{m}} \Lambda _{p00}^{m n}\,B_{p}\;, 
\label{a41} \\
\frac{d A_{0 n}^v}{d z} &=& i g\,A_{0m}^{h\,*}\,\sum_{p=0}^{\abs{m}} \Lambda _{p00}^{m n}\,B_{p}\;, 
\label{a42} \\
\frac{d B_{p}}{d z} &=& i g\,\Lambda _{p00}^{m n}\,A_{0 m}^h A_{0 n}^v\;\; (0\leq p\leq\abs{m})\;, 
\label{p0}
\end{eqnarray}
where $B_{p}$ is the amplitude of the second harmonic radial mode $p\,$. 
Note that Eq.~(\ref{p0}) imposes the relation
\begin{eqnarray}
\frac{d B_{p}}{d z} &=& \frac{\Lambda _{p00}^{m n}}{\Lambda _{000}^{m n}} \frac{d B_{0}}{d z} \;.
\end{eqnarray}
This relation implies a constrained evolution of the second harmonic mode amplitudes. Assuming null second harmonic input
($B_{p}(0) = 0$), we obtain
\begin{eqnarray}
B_{p}(z) &=& \frac{\Lambda _{p00}^{m n}}{\Lambda _{000}^{m n}} B_{0}(z) \;.
\label{bpb0}
\end{eqnarray}
By defining the rescaled amplitudes
\begin{eqnarray}
b &=& \frac{g}{\Lambda_{000}^{m n}}\,\left[\sum_{p=0}^{\abs{m}} \left(\Lambda _{p00}^{m n}\right)^2\right]\,B_{0}\;, 
\\
a_{h} &=& g\,\left[\sum_{p=0}^{\abs{m}} \left(\Lambda _{p00}^{m n}\right)^2\right]^{1/2} A_{0 m}^{h}\;, 
\\
a_{v} &=& g\,\left[\sum_{p=0}^{\abs{m}} \left(\Lambda _{p00}^{m n}\right)^2\right]^{1/2} A_{0 n}^{v}\;, 
\end{eqnarray}
we arrive at an effective three-mode coupling, governed by the same rescaled dynamical equations (\ref{rescaledprinc1}).
Therefore, the orbital angular momentum mixing of counter rotating vortices gives rise to a superposition of radial 
modes, locked in phase and amplitude as determined by Eq.(\ref{bpb0}). In some sense, this phase and amplitude locking 
among the radial modes can be considered analogous to the longitudinal mode locking achieved in pulsed laser cavities. 

For the initial condition 
$A_{0 m}^h(0) = A_{0 n}^v(0) = \sqrt{I_{0}}\,$, 
the analytical solutions for the output intensities are derived in appendix \ref{3modeanalytical}:
\begin{eqnarray}
I_{0 n}^v(z) &=& I_{0 m}^h(z) = I_{0} \,\,\mathrm{sech}^2 \left (g\,\Lambda_{000}^{m n}\,\sqrt{2I_{0}}\,z\right )\;.
\nonumber\\
J_{0}(z) &=& 2 I_{0} \,\tanh^2 \left (g\,\Lambda_{000}^{m n}\,\sqrt{2I_{0}}\,z\right ) \;,
\\
J_{p}(z) &=& \left(\frac{\Lambda _{p00}^{m n}}{\Lambda _{000}^{m n}}\right)^2\,J_{0}(z)\;.
\nonumber
\end{eqnarray}
Therefore, the second harmonic field generated by counter rotating vortices is a superposition of transverse modes 
carrying the same topological charge, but with different radial orders. Since the LG mode order is $2p+\abs{l}\,$, 
the different modes in the second harmonic field acquire different Gouy phases along propagation, making different 
near and far field images. This will be used in our experimental investigation to evidence the multimode structure 
of the second harmonic field.

\section{Experimental results}
\label{exp}

\begin{figure}
	\centering
	\includegraphics[scale=0.23]{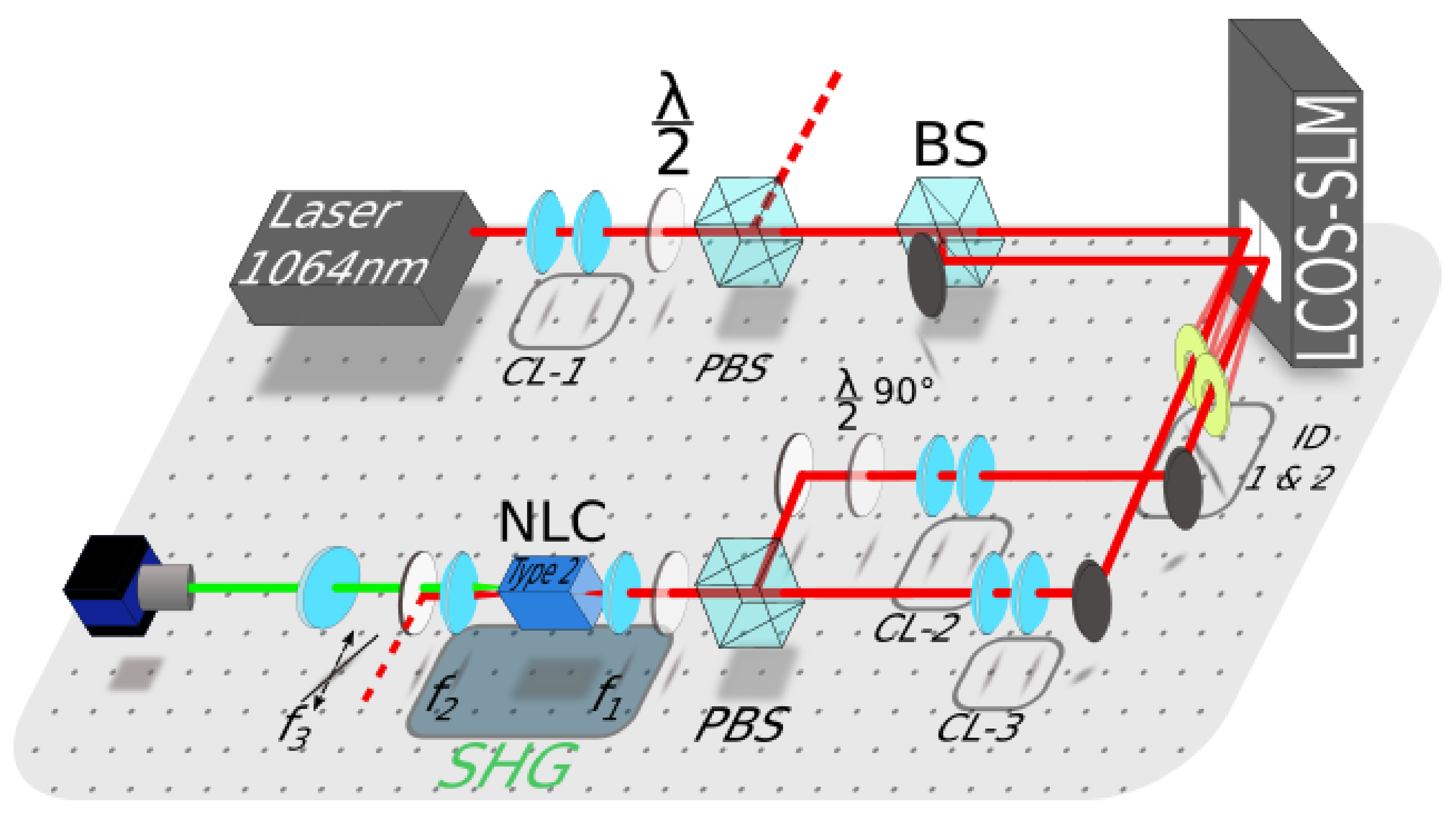}
	\caption{Experimental setup.}
	\label{fig:setup}
\end{figure}

Our experimental setup is sketched in Fig. (\ref{fig:setup}). Two beams from an infrared Nd:YAG laser (wavelength $\lambda = 1064nm$) 
impinge on two half screens of a spatial light modulator. Each half screen is computer controlled to generate an independent OAM hologram 
and produce arbitrary pairs of topological charges. The two OAM beams produced are then set with orthogonal polarizations and pass through 
pairs of collimating lenses (CL) for mode matching before being superposed on a polarizing beam splitter (PBS). 
After the PBS, the orthogonally polarized OAM beams are focused on a KTP crystal cut for type II phase match. 
The second harmonic beam at $532nm$ wavelength is separated from the fundamental beam by a spectral filter and sent to an imaging system. 

\begin{figure}
	\centering
	\includegraphics[scale=0.5]{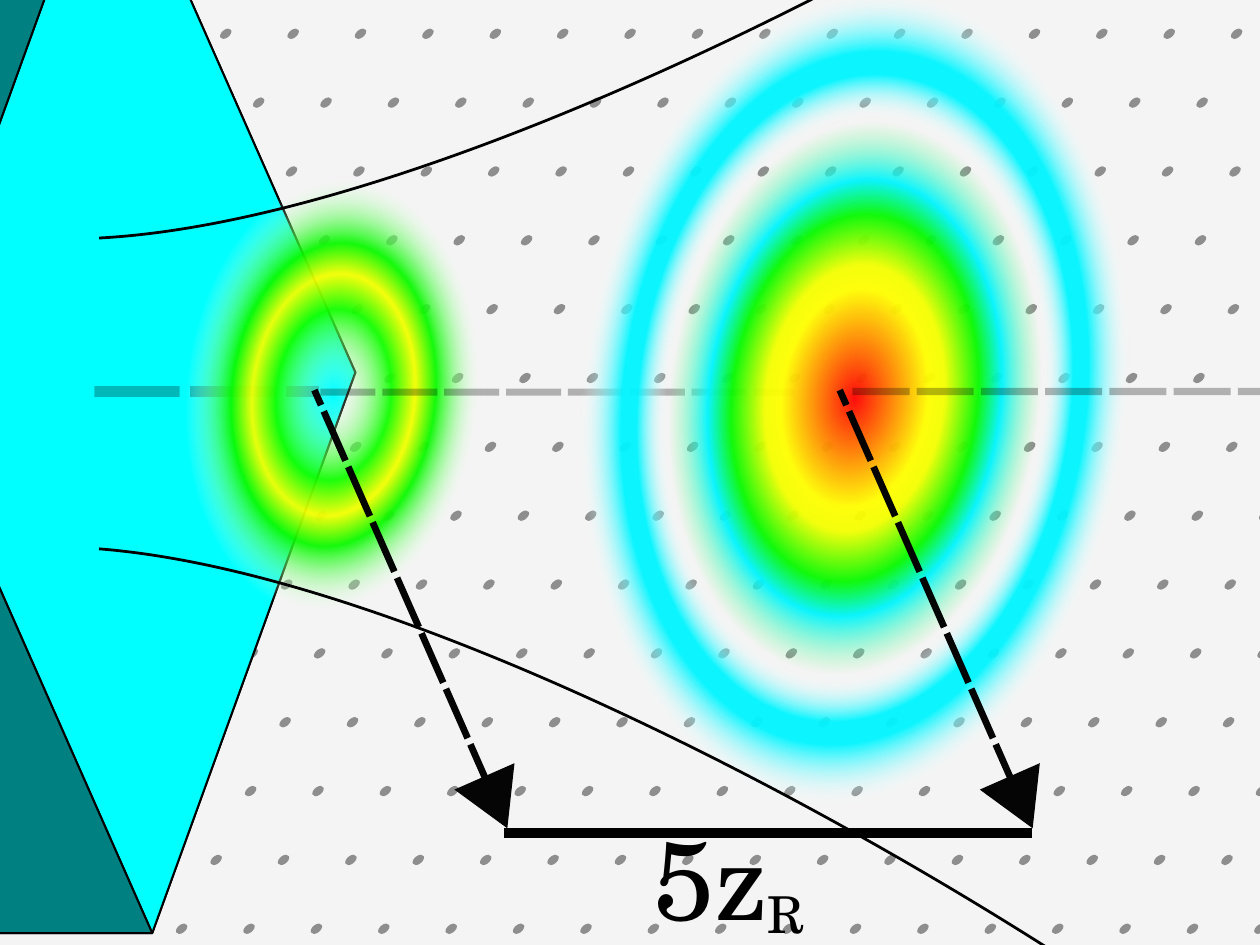}
	\caption{Sketch of the multi radial mode propagation, showing the transition between the near and far field images.}
	\label{fig:propagation}
\end{figure}

\begin{figure}
	\centering
	\includegraphics[scale=0.25]{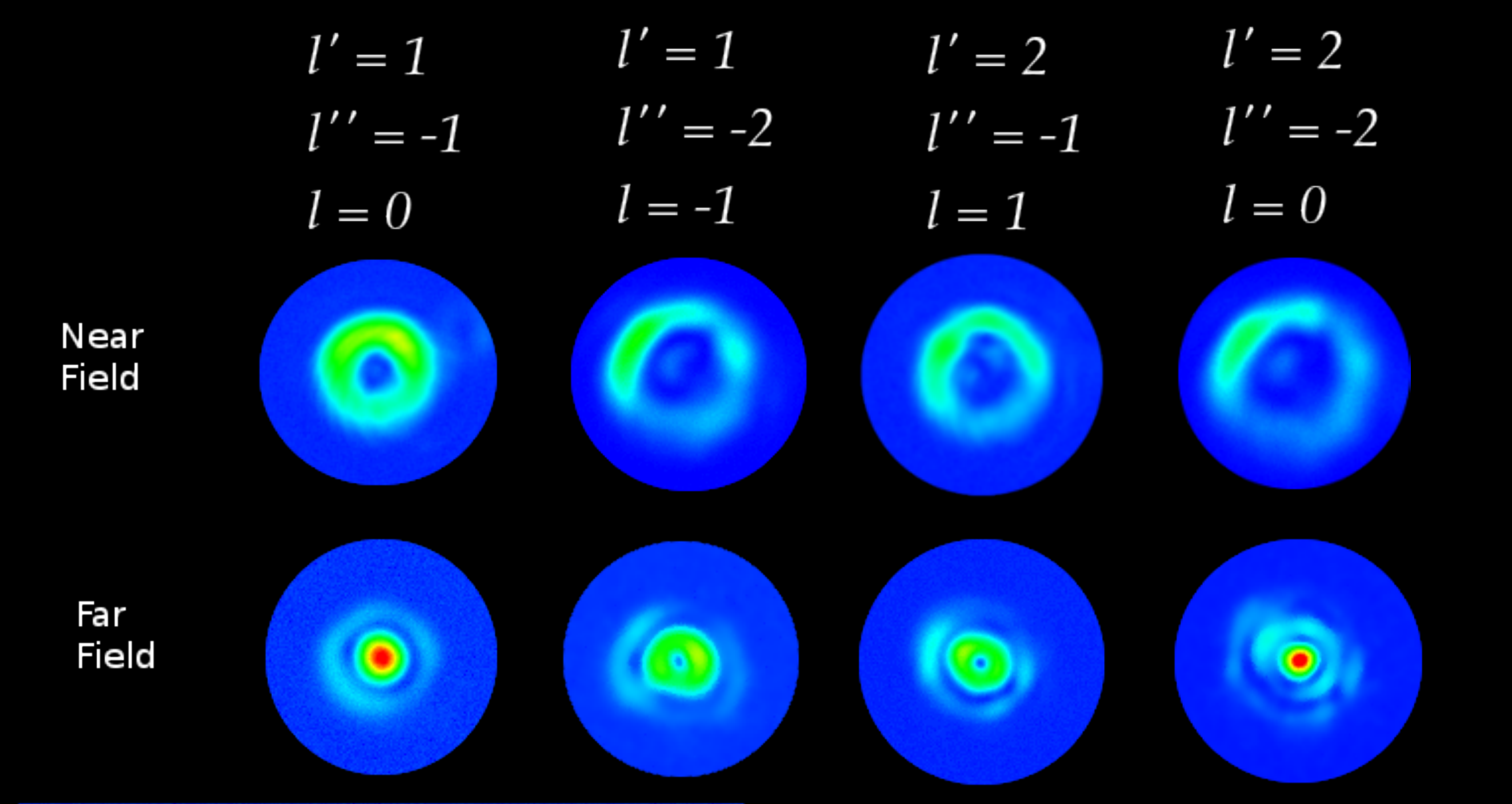}
	\includegraphics[scale=0.25]{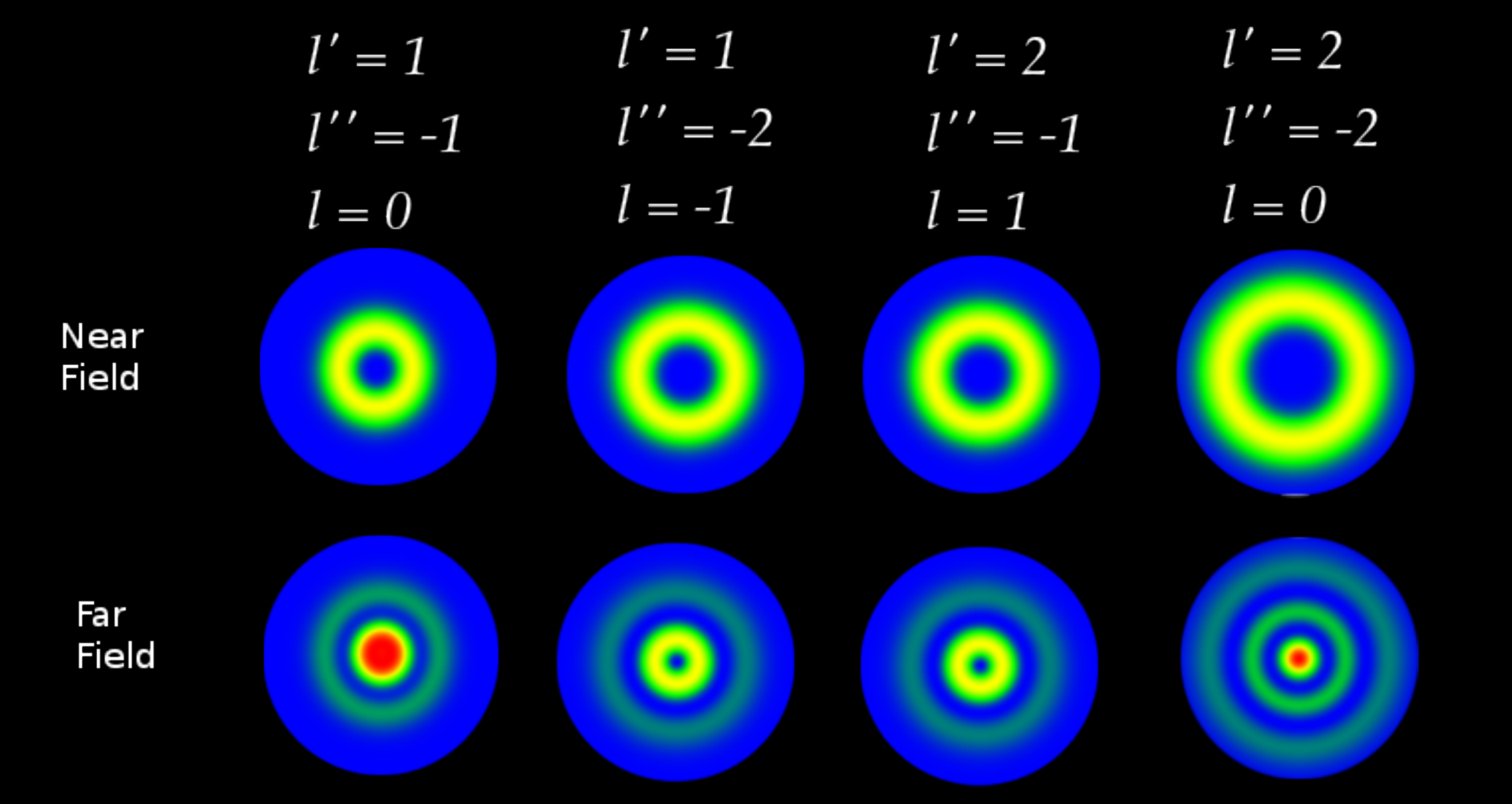}
	\caption{Experimental (top) and theoretical simulation (bottom) of the near and far field images formed by second harmonic 
	generation of counter rotating vortices.}
	\label{fig:images}
\end{figure}

In order to evidence the generation of higher radial orders experimentally, we performed intensity measurements on the second harmonic 
beam both in the near and far field regions. The superposition of different radial orders carrying the same topological charge involve 
modes acquiring different Gouy phases $(2p + \abs{l} + 1)\,\arctan(z/z_R)$ along propagation. This results in different near and far 
filed images of the second harmonic, as suggested by the propagation sketch shown in Fig. (\ref{fig:propagation}). 
The image evolution can be easily simulated and compared to the experimental images acquired with a CCD camera placed 
in different propagation regions. The corresponding images are displayed in Fig. (\ref{fig:images}). 
While the near field images only display a hollow intensity distribution, the appearance of external rings in the far field 
intensity patterns unravel the presence of higher radial orders. Moreover, the phase singularity remains present in the far 
field when there is a net OAM transferred to the second harmonic. In these cases, the phase singularity and the 
external rings coexist in the far field intensity distribution. The agreement between the experimental results and the 
simulated images is remarkable.

An intuitive picture can be envisaged for the far field structure of the second harmonic produced 
by opposite topological charges ($l^\prime = -l^{\prime\prime}$). 
In this case, the second harmonic carries no OAM, but has a hollow intensity 
distribution in the near field. Therefore, the far field structure is analogous to the diffraction pattern 
produced by a circular obstacle and the central peak can be viewed as a manifestation of the famous Poisson 
spot, which by the way also comes accompanied by external rings. Of course, in the usual Poisson spot situation, 
a sharp circular obstacle is assumed, giving rise to a virtually infinite number of external rings. However, the 
near field pattern of the second harmonic generated by opposite input charges has a smooth hollow distribution. 
Therefore, the far field pattern will exhibit a limited number of external rings, precisely equal to the absolute 
value of the mutually annihilating charges. This suggests a further method for topological charge measurement 
through the nonlinear mixing of a sample OAM beam with its mirror image.

\section{Conclusion}

We developed a detailed study of the orbital angular momentum mixing in type II second harmonic generation. 
The multimode nonlinear dynamical equations were derived and solved.
A special attention was given to the selection rules determining the transverse mode coupling in the nonlinear 
medium. While the well known OAM conservation condition is recovered, a less trivial condition was derived for 
the radial modes. The generation of higher radial orders, when counter rotating vortices are mixed, is a subtle 
effect predicted by our theoretical approach and confirmed by our experimental results. For opposite input 
charges, it allows an interesting interpretation in terms of the famous Poisson spot and can be viewed as a 
further method for OAM measurement.

Another interesting feature is the effective three-mode dynamics obtained even when higher radial orders are 
generated. These higher radial orders are phase and amplitude locked to the fundamental radial mode, producing 
a kind of transverse mode locking.

\newpage

\appendix


\section{Spatial overlap and selection rules}
\label{appoverlap}

We now calculate explicitly the overlap integrals giving rise to OAM conservation and radial order selection rules. The effective 
nonlinear coupling between modes $(p,l\,;\,2\omega)\,$, $(p',l'\,;\,\omega,h)$ and $(p'',l''\,;\,\omega,v)$ is determined by the follwing 
overlap integral:
\begin{eqnarray}
R_{pp'p''}^{ll'l''} &=& \sqrt{\frac{8}{\pi^3}}\,
\frac{\mathcal{N}_{pl}\mathcal{N}_{p'l'}\mathcal{N}_{p''l''}}{w_h\,w_v\,w_{2\omega}}
\int_{0}^{\infty }  r\,dr\,\,
\frac{\left ( \sqrt{2} r\right )^{\abs{l} + \abs{l'} + \abs{l''}}}{w_{2\omega}^{\abs{l}} w_{h}^{\abs{l'}} w_{v}^{\abs{l''}}}\,
\nonumber \\
&\times&  
L_{p}^{\abs{l}}\left ( \frac{2 r^2}{w^2_{2\omega}} \right )
L_{p'}^{\abs{l'}}\left ( \frac{2 r^2}{w^2_{h}} \right )
L_{p''}^{\abs{l''}}\left ( \frac{2 r^2}{w^2_{v}} \right ) 
\nonumber \\
&\times&  
e^{-\left(\frac{r^2}{w_{2\omega}^{\,2}}+\frac{r^2}{w_{h}^{\,2}}+\frac{r^2}{w_{v}^{\,2}}\right)} 
\int_{0}^{2\pi } d\theta\,\,e^{i (l-l'-l'')\theta}\;.
\end{eqnarray}
Note that the phase ($k_{h} + k_{v} = k_{2\,\omega}$) and wavefront ($z_{R_h} = z_{R_v} = z_{R_{2\omega}}$) 
match conditions allow the cancellation of the curved wavefront contributions and imply the following 
relationship between the mode widths
\begin{equation}
\frac{1}{w_{2\omega}^{\,2}} = \frac{1}{w_{h}^{\,2}} + \frac{1}{w_{v}^{\,2}}\;.
\end{equation}
Moreover, the angular integral provides the OAM selection rule $l=l'+l''\,$, so that the overlap integral can be written as 
\begin{eqnarray}
& & R_{pp'p''}^{ll'l''} = \sqrt{\frac{2}{\pi }}\,\delta _{l,l'+l''}\,
\frac{\mathcal{N}_{pl}\mathcal{N}_{p'l'}\mathcal{N}_{p''l''}}{w_{2\omega}}\,
\sqrt{\xi_h^{\abs{l'}+1}\,\xi_v^{\abs{l''}+1}}
\nonumber \\
&\times& \int_{0}^{\infty } x^{\frac{\left | l \right | + \left | l' \right | + \left | l'' \right |}{2}} 
L_p^{\left| l \right|} \left(x \right ) L_{p'}^{\left | l' \right |}\left ( \xi_h\,x \right ) 
L_{p''}^{\left | l'' \right |}\left ( \xi_v\,x \right ) 
e^{-x} \, dx \;,
\nonumber\\
\label{Rcobre}
\end{eqnarray}
where we defined
\begin{eqnarray}
x &=& \frac{2 r^2}{w^2_{2\omega}}\;,
\nonumber\\
\xi_j &=& \left(\frac{w_{2\omega}}{w_{j}}\right)^2\;.
\end{eqnarray}
The fundamental modes overlap is
\begin{equation}
R_{000}^{000} = \sqrt{\frac{2}{\pi }}\,\frac{\sqrt{\xi_h\,\xi_v}}{w_{2\omega}}\;.
\end{equation}
The overlap integral normalized to the fundamental modes overlap becomes
\begin{eqnarray}
& &\Lambda_{pp'p''}^{ll'l''} = 
\delta _{l,l'+l''}\,\,
\mathcal{N}_{pl}\,\mathcal{N}_{p'l'}\,\mathcal{N}_{p''l''}\,\,
\sqrt{\xi_h^{\abs{l'}}\,\xi_v^{\abs{l''}}} 
\label{Arecobre}\\
&\times& \int_{0}^{\infty } x^{\frac{\left | l \right | + \left | l' \right | + \left | l'' \right |}{2}} 
L_p^{\left| l \right|} \left(x \right ) L_{p'}^{\left | l' \right |}\left ( \xi_h\,x \right ) 
L_{p''}^{\left | l'' \right |}\left ( \xi_v\,x \right ) 
e^{-x} \, dx \;.
\nonumber
\end{eqnarray}
We will be interested in the generation of higher radial orders from zero order input beams, so we 
assume that the incoming modes have no radial structure and restrict our analysis to $p'=p''=0\,$. 
By using $L_0^k(x)=1$ and making $P = \min(\abs{l'},\abs{l''}) = (\abs{l'} + \abs{l''} - \abs{l'+ l''})/2\,$, 
the spatial overlap integral assumes the following simplified form
\begin{eqnarray}
\Lambda_{p00}^{l'l''} &=& 
\sqrt{\frac{\xi_h^{\abs{l'}}\,\xi_v^{\abs{l''}}\,p!}
	{\left ( p + \left | l'+ l'' \right | \right )! \left | l' \right |! \left | l'' \right |!}}\,\, 
\nonumber \\
&\times& 
\int_{0}^{\infty } x^{\left | l' + l'' \right |}\,x^{P} 
L_p^{\left| l' + l'' \right|} \left(x \right ) e^{-x} \, dx \;,
\label{Arecobre2}
\end{eqnarray}
where we have omitted the superfluous index $l\,$. 
From the recurrence relations for the generalized Laguerre polynomials, one readily derives a useful expansion for the monomial 
\begin{eqnarray}
x^P &=& \sum_{m=0}^{P} \frac{(-1)^m \, P! \, (k+P)!}{(P-m)! \, (k+m)!}\,L_m^k (x)\;,
\end{eqnarray}
which can be used together with the orthogonality relations, 
\begin{equation}
\int_{0}^{\infty } x^{k} L_{n}^{k}\left ( x \right ) L_{m}^{k}\left ( x \right ) e^{-x} dx = \frac{\left ( n + k \right )!}{n!} \delta _{nm}\;,
\label{Lnorma}
\end{equation}
to derive the final result
\begin{eqnarray}
\Lambda _{p00}^{l'l''} &=& \left\{
\begin{matrix}
\frac{(-1)^p\,\,\sqrt{\xi_h^{\abs{l'}}\,\xi_v^{\abs{l''}}}\,P!\,\left(P+\left | l' + l'' \right |\right)!}{\left(P - p\right)!\,
\sqrt{p!\,\left(p + \left | l' + l'' \right |\right)!\,\left | l' \right |!\,\left | l'' \right |! }} & \left(p\leq P\right)\;,
\\
0 & \left(p>P\right)\;.
\end{matrix} 
\right.
\nonumber\\
\label{Arecobrefinal}
\end{eqnarray}

Now, two different situations need to be treated separately, depending on the relative helicities of the incoming modes.
They lead to the radial selection rules for the second harmonic field.

\begin{itemize}
\item \textbf{Co-rotating vortices:} {$l'\cdot l'' \geq 0$}

This corresponds to $P=0\,$, giving
\begin{eqnarray}
\Lambda _{p00}^{l'l''} &=& 
\left\{
\begin{matrix}
\sqrt{\frac{\xi_h^{\abs{l'}}\,\xi_v^{\abs{l''}}\,\left ( \left | l' \right | + 
		\left | l'' \right | \right )!}{\left | l' \right |! \left | l'' \right |! }} & \left(p = 0\right)\;,
\\
0 & \left(p > 0\right)\;.
\end{matrix} 
\right.
\label{Arecobreco}
\end{eqnarray}
No higher radial orders are generated in the second harmonic field in this case.

\item \textbf{Counter-rotating vortices:} $l'\cdot l'' < 0$ 

This corresponds to $P=\mathrm{min}(\left | l' \right |,\left | l'' \right |)\,$, giving
\begin{eqnarray}
\Lambda _{p00}^{l'l''} &=& \left\{
\begin{matrix}
\frac{(-1)^p}{\left(P - p\right)!}\,
\sqrt{\frac{\xi_h^{\abs{l'}}\,\xi_v^{\abs{l''}}\,\abs{l'}!\,\abs{l''}!}{p!\,
		\left(p + \abs{l' + l''}\right)!\,}} & \left(p\leq P\right)\;,
\\
0 & \left(p > P\right)\;.
\end{matrix} 
\right.
\nonumber\\
\label{Arecobrecounter}
\end{eqnarray}
Higher radial orders are generated in the second harmonic field up to the 
minimum value between $\abs{l'}$ and $\abs{l''}\,$.

\end{itemize}


\newpage

\section{Multimode Manley-Rowe relations}
\label{apmanley}

In order to derive a set of multimode Manley-Rowe relations, the mode amplitudes are decomposed in terms of  
phase and intensity variables as
\begin{eqnarray}
A_{pl}^i &=& \sqrt{I_{pl}^i} e^{i \phi_{pl}^i} \;,
\nonumber\\
B_{pl} &=& \sqrt{J_{pl}} e^{i \psi _{pl}}\;.
\end{eqnarray}
In terms of these variables, the dynamical equations become
\begin{eqnarray}
& &\left (\frac{d \sqrt{I_{p'l'}^h}}{d z} + i \sqrt{I_{p'l'}^h}\,\frac{d \phi_{p'l'}^h}{d z}  \right ) 
e^{i \phi_{p'l'}^h} 
\\
&=&i g \sum_{pl}\sum_{p''l''}\Lambda _{pp'p''}^{ll'l''} \sqrt{J_{pl} I_{p''l''}^v}\,e^{i \left (\psi_{pl} - \phi_{p''l''}^v \right )}\;,
\nonumber\\
& &\left (\frac{d \sqrt{I_{p''l''}^v}}{d z} + i \sqrt{I_{p''l''}^v}\,\frac{d \phi_{p''l''}^v}{d z}  \right ) 
e^{i \phi_{p''l''}^v} 
\label{phase-intensity}\\
&=& i g \sum_{pl}\sum_{p'l'}\Lambda _{pp'p''}^{ll'l''} \sqrt{J_{pl} I_{p'l'}^h}\,e^{i \left (\psi_{pl} - \phi_{p'l'}^h \right )}\;,
\nonumber\\
& &\left (\frac{d \sqrt{J_{pl}}}{d z} + i \sqrt{J_{pl}}\,\frac{d \psi_{pl}}{d z}  \right ) 
e^{i \psi_{pl}} 
\\
&=& i g \sum_{p'l'}\sum_{p''l''}\Lambda _{pp'p''}^{ll'l''} \sqrt{I_{p'l'}^h I_{p''l''}^v}\,e^{i \left (\phi_{p'l'}^h + \phi_{p''l''}^v
	\right )}\;.
\nonumber
\end{eqnarray}
Taking the real and imaginary parts, we arrive at the phase-intensity dynamical equations
\begin{eqnarray}
\frac{d I_{p'l'}^h}{d z} &=& 2 g \sum_{pl,p''l''}\Lambda _{pp'p''}^{ll'l''} \sqrt{J_{pl} I_{p'l'}^h I_{p''l''}^v} 
\sin\Delta_{pp'p''}^{ll'l''}\;,
\nonumber\\
\frac{d I_{p''l''}^v}{d z} &=& 2 g \sum_{pl,p'l'}\Lambda _{pp'p''}^{ll'l''} \sqrt{J_{pl} I_{p'l'}^h I_{p''l''}^v} 
\sin\Delta_{pp'p''}^{ll'l''} \;,
\nonumber\\
\frac{d J_{pl}}{d z} &=& - 2 g \sum_{p'l',p''l''}\Lambda _{pp'p''}^{ll'l''} \sqrt{J_{pl} I_{p'l'}^h I_{p''l''}^v} 
\sin\Delta_{pp'p''}^{ll'l''} \;,
\nonumber\\
\\
I_{p'l'}^h \frac{d \phi_{p'l'}^h}{d z} &=& g \sum_{pl,p''l''}\Lambda _{pp'p''}^{ll'l''} \sqrt{J_{pl} I_{p'l'}^h I_{p''l''}^v} 
\cos\Delta_{pp'p''}^{ll'l''} \;,
\nonumber\\
I_{p''l''}^v \frac{d \phi_{p''l''}^v}{d z} &=& g \sum_{pl,p'l'}\Lambda _{pp'p''}^{ll'l''} \sqrt{J_{pl} I_{p'l'}^h I_{p''l''}^v} 
\cos\Delta_{pp'p''}^{ll'l''} \;,
\nonumber\\
J_{pl} \frac{d \psi_{pl}}{d z} &=& g \sum_{p'l',p''l''}\Lambda _{pp'p''}^{ll'l''} \sqrt{J_{pl} I_{p'l'}^h I_{p''l''}^v} 
\cos\Delta_{pp'p''}^{ll'l''} \;,
\nonumber\\
\end{eqnarray}
where $\Delta_{pp'p''}^{ll'l''} = \phi_{p'l'}^h + \phi_{p''l''}^v - \psi_{pl}\,$.
From these equations of motion, we derive the conservation laws given by generalized Manley-Rowe relations.
The first relation is derived by adding up the intensity equations of motion for all Laguerre-Gaussian components of 
the down converted fields, giving
\begin{eqnarray}
\frac{d }{d z} \sum_{p\,l} \left(I_{p\,l}^v - I_{p\,l}^h \right) = 0\;,
\label{Iv-Ih}
\end{eqnarray}
which simply states that the total intensity difference between signal and idler fields remains constant along the 
nonlinear process.

The second relation is derived by adding up the intensity equations of motion for the Laguerre-Gaussian components of 
the second harmonic field and comparing with the previous result obtained for the down converted fields, what gives 
the following relation 
\begin{equation}
\frac{d }{d z} \sum_{p\,l} \left( J_{p\,l} + I_{p\,l}^h \right ) = 
\frac{d }{d z} \sum_{p\,l} \left( J_{p\,l} + I_{p\,l}^v \right ) = 0 \;.
\label{J+I}
\end{equation}
These conservation laws help identifying the natural integration constants of the nonlinear dynamical equations. 
They will be used in the derivation of their analytical solutions.

\newpage

\section{Three-mode analytical solution}
\label{3modeanalytical}

We derive the analytical solution for the three-mode evolution in second-harmonic generation as described by 
the rescaled nonlinear equations
\begin{eqnarray}
\frac{d a_{h}}{d z} &=& i\,b\,a_{v}^{*}\;, 
\nonumber \\
\frac{d a_{v}}{d z} &=& i\,b\,a_{h}^{*}\;,
\label{eqav} \\
\frac{d b}{d z} &=& i\,a_{h}\,a_{v}\;.
\nonumber
\end{eqnarray}
Let us define the phase-intensity variables according to
\begin{eqnarray}
a_{h} &=& \sqrt{\bar{I}_{h}}\,e^{i\phi_h}\;, 
\label{eqah} 
\nonumber\\
a_{v} &=& \sqrt{\bar{I}_{v}}\,e^{i\phi_v}\;, 
\\
b &=& \sqrt{\bar{J}}\,e^{i\psi}\;, 
\nonumber
\end{eqnarray}

The phase-intensity dynamical equations are of the general form
\begin{eqnarray}
\frac{d \bar{I}_{h}}{d z} &=& 2 \,\sqrt{\bar{J} \bar{I}_{h} \bar{I}_{v}}\, 
\sin \left (\phi_{h} + \phi_{v} - \psi \right )\;,
\nonumber\\
\frac{d \bar{I}_{v}}{d z} &=& 2 \,\sqrt{\bar{J} \bar{I}_{h} \bar{I}_{v}}\, 
\sin \left (\phi_{h} + \phi_{v} - \psi \right )\;,
\nonumber\\
\frac{d \bar{J}}{d z} &=& - 2 \,\sqrt{\bar{J} \bar{I}_{h} \bar{I}_{v}}\, 
\sin \left (\phi_{h} + \phi_{v} - \psi \right )\;,
\nonumber\\
\bar{I}_{h}\,\frac{d \phi_{h}}{d z} &=& \sqrt{\bar{J} \bar{I}_{h} \bar{I}_{v}}\, 
\cos\left (\phi_{h} + \phi_{v} - \psi \right )\;,
\label{phase-int}\\
\bar{I}_{v}\,\frac{d \phi_{v}}{d z} &=& \sqrt{\bar{J} \bar{I}_{h} \bar{I}_{v}}\, 
\cos\left (\phi_{h} + \phi_{v} - \psi \right )\;,
\nonumber\\
\bar{J}\,\frac{d \psi}{d z} &=& \sqrt{\bar{J} \bar{I}_{h} \bar{I}_{v}}\, 
\cos\left (\phi_{h} + \phi_{v} - \psi \right )\;.
\nonumber
\end{eqnarray}
From these phase-intensity equations it can be straightforwardly demonstrated that 
\begin{eqnarray}
\frac{d}{dz}\left(\bar{J}\,\frac{d \psi}{d z}\right) = 
\frac{d}{dz}\left(\bar{I}_{h}\frac{d \phi_{h}}{d z}\right) =  
\frac{d}{dz}\left(\bar{I}_{v}\frac{d \phi_{v}}{d z}\right) = 0\;,
\nonumber\\
\end{eqnarray}
giving the following conserved quantity
\begin{eqnarray}
\sqrt{\bar{I}_{h}(z) \bar{I}_{v}(z) \bar{J}(z)} \,\cos\Phi (z) = 
\sqrt{\bar{I}_{h}(0) \bar{I}_{v}(0) \bar{J}(0)} \,\cos\Phi (0)\;,
\nonumber\\
\end{eqnarray}
where $\Phi(z)\equiv\phi_{h}(z) + \phi_{v}(z) - \psi(z)\,$. If the second harmonic field is not seeded, 
then $\bar{J}(0)=0$ and this conserved quantity imposes $\cos\Phi(z) = 0$ for non vanishing solutions for $\bar{J}(z)\,$. 
In this case, one immediately sees from the phase equations that $\phi_h\,$, $\phi_v$ and $\psi$ are 
stationary and must fulfill $\Phi (z) = (2n + 1)\,\pi/2\,$.

In order to derive the analytical solution for the three-mode evolution, we assume that the input fields 
have equal intensities $\bar{I}_{v}(0) = \bar{I}_{h}(0) = \bar{I}_0\,$. From the Manley-Rowe relations one has 
\begin{eqnarray}
\bar{I}_{v}(z) &=& \bar{I}_{h}(z)\;, \label{MRI1igualI213} \\
\bar{J}(z) &=& 2 \left[ \bar{I}_0 - \bar{I}_{h}(z) \right]\;, \label{MRI1igualI223}
\end{eqnarray}
so that the dynamical equation for $\bar{I}_{h}$ assumes the simple form
\begin{eqnarray}
\frac{d \bar{I}_{h}}{d z} = 2 \sqrt{2(\bar{I}_0-\bar{I}_{h})}\,\bar{I}_{h}\;,
\label{acabando}
\end{eqnarray}
which can be readily solved, giving the well known hyperbolic solutions for the interacting field intensities  
\begin{eqnarray}
\bar{I}_{h}(z) &=& \bar{I}_{v}(z) = \bar{I}_0 \,\,\mathrm{sech}^2 \left (\sqrt{2\bar{I}_0} \, z \right )\;,
\nonumber\\
\bar{J}(z) &=& 2 \bar{I}_0 \tanh^2 \left (\sqrt{2\bar{I}_0} \, z \right )\;.
\end{eqnarray}


\begin{thebibliography}{100}


\bibitem{oam25}
M. Padgett,
Opt. Exp. \textbf{25}, 11265 (2017).



\bibitem{poincare}
M. J. Padgett, J. Courtial,
Opt. Lett. \textbf{24}, 430 (1999).

\bibitem{imagepol}
D. P. Caetano, P. H. Souto Ribeiro, J. T. C. Pardal, and A. Z. Khoury, 
Phys. Rev. A {\bf 68}, 023805 (2003).

\bibitem{topoluff}
C. E. R. Souza, J. A. O. Huguenin, P. Milman, and A. Z. Khoury, 
Phys. Rev. Lett. {\bf 99}, 160401 (2007). 

\bibitem{quplate}
E. Nagali, F. Sciarrino, F. De Martini, L. Marrucci, B. Piccirillo, E. Karimi, and E. Santamato.
Phys. Rev. Lett. \textbf{103}, 013601 (2009). 

\bibitem{simon}
B. N. Simon, S. Simon, F. Gori, M. Santarsiero, R. Borghi, N. Mukunda, and R. Simon,
Phys. Rev. Lett. \textbf{104}, 023901 (2010).

\bibitem{aiello1}
A. Holleczek, A. Aiello, C. Gabriel, C. Marquardt, andG.  Leuchs, 
Opt. Express. \textbf{19}, 9714 (2011).

\bibitem{cardano}
F. Cardano, E. Karimi, S. Slussarenko, L. Marrucci, C. de Lisio, and E. Santamato, 
Appl. Opt. \textbf{51}, C1-C6 (2012).

\bibitem{milione}
G. Milione, A. Dudley, T. A. Nguyen, K. Chakraborty, E. Karimi, A. Forbes, and R. R. Alfano, 
J. Opt. \textbf{17}, 035617 (2015).

\bibitem{aiello2}
A.  Aiello,  F.  T\"oppel,  C.  Marquardt,  E.  Giacobino,  G. Leuchs,
New J. Phys. \textbf{17}, 043024 (2015).

\bibitem{structured-waves}
J. Harris, V. Grillo, E. Mafakheri, G. C. Gazzadi, S. Frabboni, R. W. Boyd, E. Karimi,
Nat. Phys. \textbf{11}, 629 (2015). 


\bibitem{milione2}
G. Milione, T. A. Nguyen, J. Leach, D. A. Nolan, and R. R. Alfano,
Opt. Lett. \textbf{40}, 4887 (2015).

\bibitem{network}
Zhi-Yuan Zhou, Yan Li, Dong-Sheng Ding, Wei Zhang, Shuai Shi, Bao-Sen Shi,
Opt. Exp. \textbf{23}, 18435 (2015).


\bibitem{chen2009} 
L. Chen and W. She, 
Phys. Rev. A {\bf 80}, 063831 (2009).

\bibitem{barreiro2010}
J. T. Barreiro, T.-C. Wei, and P. G. Kwiat, 
Phys. Rev. Lett. {\bf 105},  030407 (2010).

\bibitem{khourymilman}
A. Z. Khoury and P. Milman,
Phys. Rev. A \textbf{83}, 060301(R) (2011). 

\bibitem{rafsanjani}
S. M. H. Rafsanjani, M. Mirhosseini, O. S. Maga\~na-Loaiza, R. W. Boyd,
Phys. Rev. A \textbf{92}, 023827 (2015). 

\bibitem{remote-state-boyd}
M. Erhard, H. Qassim, H. Mand, E. Karimi, R. W. Boyd,
Phy. Rev. A \textbf{92}, 022321 (2015).

\bibitem{szameit}
D. Guzman-Silva, R. Br\"uning, F. Zimmermann, C. Vetter, M. Gr\"afe, M. Heinrich, 
S. Nolte, M. Duparr\'e, A. Aiello, M. Ornigotti, A. Szameit,
Laser Photonics Rev. \textbf{10}, 317 (2016).

\bibitem{teleportUFF}
B. Pinheiro da Silva, M. Astigarreta Leal, C. E. R. Souza, E. F. Galv\~ao and A. Z. Khoury, 
J. Phys. B, \textbf{49}, 055501 (2016).



\bibitem{cryptosteve}
L. Aolita and S. P. Walborn. 
Phys. Rev. Lett. \textbf{98}, 100501 (2007). 

\bibitem{cryptouff}
C. E. R. Souza, C. V. S. Borges, A. Z. Khoury, J. A. O. Huguenin, L. Aolita, and S. P. Walborn. 
Phys. Rev A \textbf{77}, 032345 (2008).

\bibitem{cryptolorenzo}
V. D'Ambrosio, E. Nagali, S. P. Walborn, L. Aolita, S. Slussarenko, L. Marrucci, and F. Sciarrino
Nat. Comm. \textbf{3}, 961 (2012). 



\bibitem{cnotsteve}
A. N. de Oliveira, S. P. Walborn,  and C. H. Monken. 
J. Opt. B: Quantum Semiclassic. Opt. \textbf{7}, 288 (2005). 

\bibitem{cnotuff}
C. E. R. Souza and A. Z. Khoury,
Opt. Express {\bf 18}, 9207 (2010). 


\bibitem{josabuff}
A. R. C. Pinheiro, C. E. R. Souza, D. P. Caetano, J. A. O. Huguenin, A. G. M. Schmidt and A. Z. Khoury,
J. Opt. Soc. Am. B \textbf{30}, 3210 (2013). 

\bibitem{environ}
M. Hor-Meyll, A. Auyuanet, C. V. S. Borges, A. Arag\~ao, J. A. O. Huguenin, A. Z. Khoury and L. Davidovich, 
Phys. Rev. A \textbf{80}, 042327 (2009). 


\bibitem{aiello3}
F. T\"oppel, A. Aiello, C. Marquardt, E. Giacobino and G.Leuchs, 
New. J. Phys. \textbf{16}, 073019 (2014).

\bibitem{forbes}
M. McLaren, T. Konrad, and A. Forbes, 
Phys. Rev. A \textbf{92}, 023833 (2015).

\bibitem{forbes3}
B. Ndagano, B. Perez-Garcia, F. S. Roux, M. McLaren, C. Rosales-Guzman, Y. Zhang, O. Mouane, 
R. I. Hernandez-Aranda, T. Konrad, A. Forbes,
Nat. Phys. \textbf{13}, 397 (2017).



\bibitem{bell2}
C. V. S. Borges, M. Hor-Meyll, J. A. O. Huguenin, and A. Z. Khoury,
Phys. Rev. A {\bf 82}, 033833 (2010).  

\bibitem{chen}
L. Chen and W. She, 
J. Opt. Soc. Am. B \textbf{27}, A7 (2010).

\bibitem{bellpadgett}
E. Karimi, J. Leach, S. Slussarenko, B. Piccirillo, L. Marrucci, 
L. X. Chen, W. L. She, S. Franke-Arnold, M. J. Padgett, E. Santamato, 
Phys. Rev. A \textbf{82}, 022115 (2010). 

\bibitem{kagalwala}
K. H. Kagalwala, G. Di Giuseppe, A. F. Abouraddy, and B. E.A. Saleh, 
Nat. Photon. \textbf{7}, 72 (2013).

\bibitem{eberly}
X. F. Qian and J. H. Eberly, 
Opt. Lett. \textbf{36}, 4110 (2011).

\bibitem{eberly2}
X. F. Qian, B. Little, J. C. Howell and J. H. Eberly, 
Optica \textbf{2}, 611 (2015).

\bibitem{tripartite}
W. F. Balthazar, C. E. R. Souza, D. P. Caetano, E. F. Galv\~ao, J. A. O. Huguenin, A. Z. Khoury,
Opt. Lett. \textbf{41}, 5797 (2016).



\bibitem{SHG} 
J. Courtial, K. Dholakia, L. Allen, and M. J. Padgett,
Phys. Rev. A \textbf{56}, 4193 (1997).

\bibitem{SUM} 
A. Berzanskis, A. Matijosius, A. Piskarskas, V. Smilgevicius and A. Stabinis,
Opt. Commun. \textbf{150}, 372-380 (1998).

\bibitem{OAM-SHG-Li}
Si-Min Li, Ling-Jun Kong, Zhi-Cheng Ren, Yongnan Li, Chenghou Tu, Hui-Tian Wang,
Phys. Rev. A \textbf{88}, 035801 (2013).

\bibitem{OAM-SHG-UFF}
W. T. Buono, L. F. C. Moraes, J. A. O. Huguenin, C. E. R. Souza and A. Z. Khoury,
New Journal of Physics \textbf{16}, 093041 (2014).

\bibitem{sum-freq}
Yan Li, Zhi-Yuan Zhou, Dong-Sheng Ding, Bao-Sen Shi,
J. Opt. Soc. Am. B \textbf{32}, 407 (2015).

\bibitem{algebra}
A. A. Zhdanova, M. Shutova, A. Bahari, M. Zhi, A. V. Sokolov,
Opt. Exp. \textbf{23}, 34109 (2015).

\bibitem{twisted-nonlinear} 
Yan Li, Zhi-Yuan Zhou, Dong-Sheng Ding, Bao-Sen Shi,
J. Mod. Opt. \textbf{63}, 2271 (2016). 

\bibitem{beam-prop}
T. Yusufu, Y. Sasaki, S. Araki, K. Miyamoto, T. Omatsu,
App. Opt. \textbf{55}, 5263 (2016).

\bibitem{shg-plasmonic}
Xiaoyan Y. Z. Xiong, Ahmed Al-Jarro, Li Jun Jiang, Nicolae C. Panoiu, Wei E. I. Sha,
Phys. Rev. B \textbf{95}, 165432 (2017).



\bibitem{PDC1}
J. Arlt, K. Dholakia, L. Allen, and M. J. Padgett, 
Phys. Rev. A \textbf{59}, 3950 (1999).

\bibitem{zeilinger} 
A. Mair, A. Vaziri, G. Weihs and A. Zeilinger, 
Nature \textbf{412}, 313 (2001).

\bibitem{PDCUFF}
D. P. Caetano, M. P. Almeida, P. H. Souto Ribeiro, J. A. O. Huguenin, B. Coutinho dos Santos, and A. Z. Khoury,
Phys. Rev. A \textbf{66}, 041801 (2002).


\bibitem{Schwob}
C. Schwob, P. F. Cohadon, C. Fabre, M. A. M. Marte, H. Ritsch, A. Gatti, L. Lugiato,
Appl Phys B \textbf{66}, 685-699 (1998). 

\bibitem{OPO1}
M. Martinelli, J. A. O. Huguenin, P. Nussenzveig, A. Z. Khoury,
Phys. Rev. A \textbf{70}, 013812 (2004).	

\bibitem{OPOUFF}
B. Coutinho dos Santos, A. Z. Khoury, J. A. O. Huguenin, 
Opt. Lett. \textbf{33}, 2803 (2008).

\bibitem{OPO2}
A. Aadhi, G. K. Samanta, S. Chaitanya Kumar, and M. Ebrahim-Zadeh,
Optica \textbf{4}, 349 (2017).


\bibitem{ATO1} 
J. W. R. Tabosa and D. V. Petrov,
Phys. Rev. Lett. \textbf{83}, 4967 (1999).

\bibitem{ATO2} 
S. Barreiro, J. W. R. Tabosa,
Phys. Rev. Lett. \textbf{90}, 133001 (2003).


\bibitem{atto}
R. G\'eneaux, A. Camper, T. Auguste, O. Gobert, J. Caillat, R. Ta\"{\i}eb, T. Ruchon,
Nat. Comm. \textbf{7}, 12583 (2016).

\bibitem{high-hg}
F. Kong, C. Zhang, F. Bouchard, Z. Li, G. G. Brown, D. H. Ko, T. J. Hammond, L. Arissian, R. W. Boyd, E. Karimi, P. B. Corkum,
Nat. Comm. \textbf{8}, 14970 (2017). 


\bibitem{pol-shaping}
F. Bouchard, H. Larocque, A. M. Yao, C. Travis, I. De Leon, A. Rubano, E. Karimi, G-L. Oppo, R. W. Boyd,
Phys. Rev. Lett. \textbf{117}, 233903 (2016). 



\bibitem{radial-pdc}
Y. Zhang, F. S. Roux, M. McLaren, A. Forbes,
Phys. Rev. A \textbf{89}, 043820 (2014).


\bibitem{bloembergen}
J. A. Armstrong, N. Bloembergen, J. Ducuing, and P. S. Pershan
Phys. Rev. \textbf{127}, 1918 (1962).

\end{thebibliography}
\end{document}